\documentclass{pazha2}
\usepackage{epsf}
\usepackage{graphicx}
\usepackage{multirow}
\usepackage{color}
\definecolor{darkblue}{rgb}{0,0,0.9}

\begin{document}
\journalinfo{2018}{44}{12}{777}{845}{850}[781]
\sloppypar

\title{Single X-ray Bursts and the Model of a Spreading Layer of Accreting
Matter over the Neutron Star Surface} \year=2018 \author{
  ~ \ \
{S. A.~Grebenev}\email{grebenev@iki.rssi.ru}\addres{0},
{I. V.~Chelovekov}\addres{0}   
\addrestext{0}{Space Research Institute, Russian Academy of
  Sciences, ul. Profsoyuznaya 84/32, Moscow, 117997 Russia}
}
\shortauthor{GREBENEV, CHELOVEKOV}
\shorttitle{SINGLE X-RAY BURSTS AND THE MODEL OF A SPREADING LAYER} 

\submitted{November 6, 2017}

\begin{abstract}
\noindent
The excess of the rate of type I X-ray bursts over that expected
when the matter fallen between bursts completely burns out in a
thermonuclear explosion is explained in terms of the model of a
spreading layer of matter coming from the accretion disk over
the neutron star surface. Such excess is observed in bursters
with a high persistent luminosity, $4\times 10^{36}\ \mbox{erg
  s}^{-1}\la L_{X}\la 2\times 10^{37}\ \mbox{erg s}^{-1}$. In
this model the accreting matter settles to the stellar surface
mainly in two high-latitude ring zones. Despite the subsequent
spreading of matter over the entire star, its surface density in
these zones turns out to be higher than the average one by 2--3
orders of magnitude, which determines the predominant ignition
probability.  The multiple events whereby the flame after the
thermonuclear explosion in one ring zone (initial burst)
propagates through less dense matter to another zone and
initiates a second explosion in it (recurrent burst) make a
certain contribution to the observed excess of the burst
rate. However, the localized explosions of matter in these
zones, after which the burning in the zone rapidly dies out
without affecting other zones, make a noticeably larger
contribution to the excess of the burst rate over the expected
one.\\


\noindent
    {\bf DOI:} 10.1134/S1063773718120083\\
    
\end{abstract}

\section{INTRODUCTION}
\noindent  
Previously (Grebenev and Chelovekov 2017) we showed that the
model of a spreading layer of accreting matter over the surface
of a neutron star with a weak magnetic field (Inogamov and
Sunyaev 1999, 2010) allows the origin of multiple type I X-ray
bursts detected sometimes from the Galactic X-ray bursters to be
successfully explained. By multiple bursts we mean double or
triple ones with a recurrence time (total duration of the series
of events) $t_{r}\sim 400$--$1200$ s ($7$--$20$ min). This time
is much shorter than the characteristic accumulation time
$t_{a}$ of the critical mass needed for a thermonuclear
explosion to be initiated on the neutron star surface.

Indeed, if the matter coming from the accretion disk spreads
uniformly over the stellar surface and burns out completely
during the explosion, then $t_a\simeq 4\pi \Sigma_c R_*^2
\dot{M}^{-1}\simeq 1.7\ \Sigma_{\rm
  He}\,R_{12}^2\,\dot{M}_{17}^{-1}$ days. Here we substituted
the critical surface density of the matter
$\Sigma_c=1.7\times10^8\, \Sigma_{\rm He}\ \mbox{g cm}^{-2}$,
corresponding to explosive helium ignition (Tutukov and Ergma
1979; Ergma and Tutukov 1980; Fuijmoto et al. 1981; Hanawa and
Fujimoto 1982; Bildsten 1998). The total luminosity of the
neutron star in the period between bursts corresponding to
$\dot{M}=10^{17}\, \dot{M}_{17}\ \mbox{g s}^{-1}$ is
$L_{a}=GM_*\dot{M}/R_*\simeq 1.6\times10^{37}
(M_{1.4}\,\dot{M}_{17}/R_{12})\ \mbox{erg s}^{-1}\simeq
0.09\ (\dot{M}_{17} /R_{12}) L_{\rm ed}$. As usual,
$M_*=1.4\ M_{1.4}\ M_{\odot}$ and $R_*=12\ R_{12}\ \mbox{km}$
are typical neutron star mass and radius, $L_{\rm ed}\simeq
1.3\times10^{38} (M_*/M_{\odot})\ \mbox{erg s}^{-1}$ is the
critical Eddington luminosity, and $G$ is the gravitational
constant.
\begin{figure*}[tp]
\epsfxsize=0.95\textwidth
\epsffile{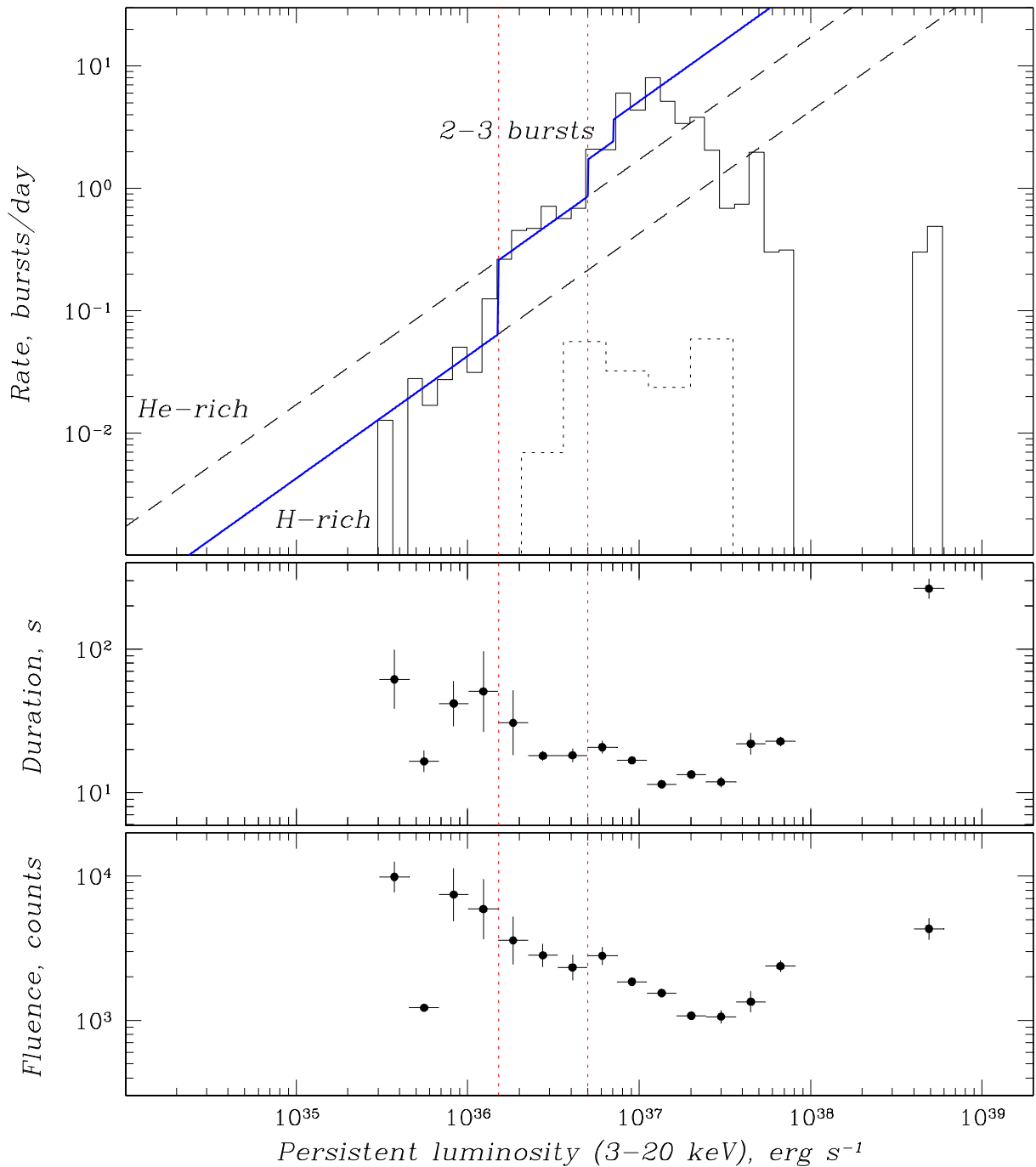}
\caption{\rm (a) Rate of X-ray bursts from bursters
  reconstructed from observations versus the accretion
  luminosity. The dashed lines indicate the predictions of the
  model of complete burning in a burst of the matter fallen to
  the neutron star surface since the previous burst. The upper
  and lower lines correspond to the different presumed helium
  abundances. The thick solid line indicates the prediction with
  allowance made for the spreading layer. (b) Mean burst
  duration versus the accretion luminosity. (c) Mean fluence in
  a burst (in counts) versus the burster luminosity.}
\end{figure*}

The seemingly unresolvable discrepancy between the times $t_r$
and $t_a$ may be explained in the model of a spreading layer of
matter over the neutron star surface. Inogamov and Sunyaev
(1999) showed that at a high ($\dot{M}\ga10^{17}\ \mbox{g
  s}^{-1}$) accretion rate the matter coming from the accretion
disk and rotating with the Keplerian velocity can not entirely
settle to the neutron star surface directly in the region of its
contact with the disk\footnote{Some fraction of the matter,
  nevertheless, must fall in precisely here to slow down the
  radial flow velocity in the disk (Grebenev and Chelovekov
  2017).}. This is hampered by the pressure of the radiation
emitted as the matter decelerates, which upsets the balance
between the gravitational and centrifugal forces. According to
Inogamov and Sunyaev (1999), the bulk of this
radiation-dominated, levitating matter is displaced in a spiral
toward the poles and settles to the stellar surface only at high
latitudes in two ring zones. Despite the subsequent spreading of
the matter already fallen to the stellar surface, its efficient
accumulation occurs in these zones (Inogamov and Sunyaev 1999).

Having noted that the conditions favorable for thermonuclear
ignition are created precisely in these ring zones, Grebenev and
Chelovekov (2017) suggested the following scenario for the
formation of multiple bursts. The explosion that began in one of
the ring zones rapidly burns out hydrogen and helium in it,
forming the first burst in the source's X-ray light curve. Then,
the thermonuclear flame slowly, with a deflagration wave speed
$v_{\rm def}\sim10-100\ \rm{m~s}^{-1}$ (Fryxell and Woosley
1982; Nozakura et al. 1984; Bildsten 1995), propagates through
matter with a lower surface density to the opposite ring zone
and initiates a new explosion there (the last burst in the
source's light curve). If a sufficient amount of matter was
accumulated in the equatorial zone, then a middle burst is
formed in the light curve when the flame front passes through
it. This event is always weaker than the first and last ones. A
triple burst is observed in this case. If there was little
matter in the equatorial zone, then a double burst is
observed. In such a way $t_r$ is the time of flame propagation
over the neutron star surface while $t_a$ is the time between
this series of events and the nearest burst not included in it.

In this paper we show that the model of a spreading layer allows
another puzzle of X-ray bursts to be explained equally
successfully --- the excess of the burst rate observed for
sources with a moderately high persistent luminosity,
$4\times10^{36}\ \mbox{erg s}^{-1}<L_X<2\times10^{37}\ \mbox{erg
  s}^{-1}$ over that predicted in the model of complete burning
of the matter fallen to the neutron star surface upon accretion
during the explosion.

\section*{OBSERVATIONS AND RESULTS}
\noindent
A catalog of type I X-ray bursts from a large number of bursters
based on the sky observations by the JEM-X and IBIS/ISGRI
telescopes onboard the INTEGRAL observatory in 2003--2015 is
presented in Chelovekov et al. (2017). This paper continues the
investigation of thermonuclear X-ray bursts with the INTEGRAL
telescopes begun in 2006 (Chelovekov et al. 2006; Chelovekov and
Grebenev 2011).  The full catalog of detected bursts is
accessible at {\tt $<$http://dlc.rsdc.rssi.ru$>$}. The large size of the
sample of bursts (2201 events) allows one to carry out their
various statistical studies, in particular, to find the
dependence of the mean rate of bursts from bursters on their
accretion luminosity (accretion rate).

This dependence is indicated in Fig.\,1a by the thin solid line
(histogram). On the whole, it is consistent with the analogous
dependence derived by Galloway et al. (2008) from RXTE data,
though even surpasses it in the number of events used. Note that
no event was recorded at luminosities $L\,(3-20\ \mbox{keV}) \la
2\times 10^{35}\ \mbox{erg s}^{-1}$. Four bursts corresponding
to a separate peak in the range of super-Eddington luminosities
were detected from the same source, GX\,17+2; they all belong to
the group of superlong bursts associated with explosive burning
of matter with a large CNO abundance (Cumming and Bildsten 2001;
in't Zand et al. 2004).

The dashed straight lines in the figure indicate the expected
change in the burst rate under the assumption that
the matter fallen to the stellar surface after the previous
burst burns out completely. The upper and lower
lines correspond to the cases of  helium and hydrogen 
burning, respectively.

According to Bildsten (1998, 2000; see also
Ergma 1983; Strohmayer and Bildsten 2006), the
following regimes of burning/explosion on the neutron
star surface are possible, depending on the local rate of
accretion $\dot{M}$:
\begin{enumerate}
  \item at $\dot{M}_{17}<0.16\ \mbox{g s}^{-1}$ (a low accretion
    rate corresponding to the luminosity $L_a=GM_*\dot{M}/R_*$ in units of
    $10^{36}\ \mbox{erg s}^{-1}$  $L_{36}<2.6$) there is
    simultaneous explosive burning of a mixture of hydrogen and
    helium caused by the explosion of hydrogen (the bursts occur
    rarely, but are characterized by a large power and
    duration);
   
  \item at $\dot{M}_{17}<0.72\ \mbox{g s}^{-1}$ (a higher
    accretion rate corresponding to $L_{36}<11.6$) hydrogen in
    the matter fallen to the neutron star surface has time to
    burn into helium in a stationary regime and, therefore, the
    explosion occurs in helium-rich matter (the bursts occur
    more frequently, but they are less powerful and shorter);

  \item at $\dot{M}_{17}<12\ \mbox{g s}^{-1}$ (an even higher
    accretion rate corresponding to $L_{36}<190$) hydrogen has
    no time to burn completely into helium and, therefore,
    helium explodes in a hydrogen-rich medium (an H/He mixture
    if, of course, the normal star in the system is not a helium
    dwarf) (the bursts turn out to be more powerful than those
    in the case of purely helium bursts);

  \item at $\dot{M}_{17}>12\ \mbox{g s}^{-1}$ ($L_{36}>190$)
    only simultaneous stationary burning of hydrogen and helium
    is possible.
\end{enumerate}
Note that the given values of $L_{36}$ relate to the total
luminosity of the source in the whole energy range, with taking
into account the emission of both the accretion disk and the
boundary or spreading layer. It is unlikely that the disk
emission contributes significantly to the X-ray flux in the
3--20 keV band.  Therefore the luminosity should be decreased by
a factor of 2 when the regimes of burst generation are
compared with Fig.\,1.

Given the above picture, the luminosity dependence of the
accretion rate observed in Fig.\,1a could be modeled as
indicated by the thick (blue) solid line. The transition from
hydrogen explosive burning to helium one is qualitatively
confirmed by the dependences of the burst duration and fluence
(in counts) on the burster continuum luminosity presented in
Figs.\,1b and 1c. Indeed, there is an obvious decrease of the
mean burst duration\footnote{Actually there are two burst
  populations observed below this threshold of luminosities,
  more or less short ones and long ones, thus the term of the
  mean burst duration is rather arbitrary.} and fluence in them
in the range $L(3-20\ \mbox{keV})\geq 1.5\times
10^{36}\ \mbox{erg s}^{-1}$, along with the step-wise increase
in the burst rate.  However, the observed dependence of the
burst rate on the luminosity (accretion rate) is consistent with
the described picture only at luminosities
$L\,(3-20\ \mbox{keV}) \la 4\times 10^{36}\ \mbox{erg
  s}^{-1}$. At higher luminosities, in the range $4\times
10^{36}\ \mbox{erg s}^{-1}\la L\,(3-20\ \mbox{keV}) \la 2\times
10^{37}\ \mbox{erg s}^{-1}$, the observed burst rate exceeds the
expected one by several times.

The enhanced burst rate in this luminosity range can be
explained in terms of the model of a spreading layer of
accreting matter over the neutron star surface.  As has been
said above, at such high luminosities the matter falls to the
stellar surface in two high-latitude ring zones and it has no
time to spread over the entire surface during the characteristic
time interval between bursts. The accumulation of matter occurs
in these zones, each of which can be responsible for its burst.

Once an explosion has occurred in one zone (for more details
see, e.g., Grebenev and Chelovekov 2017), the thermonuclear
burning can propagate with a deflagration wave speed through
less dense matter to another zone and initiate an explosion
there. In this case, the bursts form a series of close events,
multiple bursts. In Fig.\,1 all recurrent bursts in the recorded
series were removed; only the initial bursts were left.  The
removed bursts of the series are indicated in the figure by the
dotted histogram. They are not too many. The rarity of such
multiple bursts shows that the flame front by no means always
manages to reach another zone; in most cases, the burning in the
zone rapidly dies out without affecting other zones. However, as
a result, the number of bursts doubles or even triples (if an
appreciable fraction of the matter settles to the surface
directly in the equatorial zone of the neutron star, in the
place of contact of the star and the accretion disk; see
Grebenev and Chelovekov 2017) compared to the number of bursts
expected in the case of complete uniform spreading of matter
over the neutron star surface.

The thick (blue) solid line in Fig.\,1a shows how such
doubling/tripling of the number of bursts at such high
luminosities (accretion rates) of bursters allows the
observations to be explained. Note that at even higher
luminosities $L\,(3-20\ \mbox{keV}) \ga 2\times
10^{37}\ \mbox{erg s}^{-1}$ the burst rate expectedly (see Lewin
et al. 1993; Bildsten 1998) drops in view of the gradual
transition of thermonuclear burning to a continuous regime or a
noticeable rise in the radiation pressure and, along with it,
the critical surface density of matter needed for an explosion
to be initiated on the neutron star. However, this is already a
completely different problem.

The possibility of burst generation at high latitudes (near the
poles of the neutron star) at rather high accretion rates
(corresponding to $0.19 L_{\rm ed}< L_a < 0.34 L_{\rm ed}$) was
theoretically predicted previously by Cooper and Narayan
(2007). They did not consider spreading the matter over the
neutron star surface but investigated stability of thermonuclear
burning at different latitudes. In particular, they showed that
the probability of matter ignition is maximal near the equator
when the rate corresponds to $L_a< 0.16 L_{\rm ed}$ (see also
Spitkovskiy et al. 2002). Thus the bursts in Fig.\,1 at low
luminosities can actually arise at the near-equatorial region
(note that at such luminosities the matter settles to the
surface near this region and has enough time for spreading over
the broader region). While the accretion rate increases till the
level corresponding to $L_a\sim 0.19 L_{\rm ed}$, the
probability of explosion at high latitudes increases and becomes
dominant at $L_a> 0.19 L_{\rm ed}$. Moreover, thermonuclear
burning at low latitudes (near the equator) is stabilized and
bursts could not arise at all in this region at the rates
corresponding to $L_a> 0.30 L_{\rm ed}$. The burning becomes
stable over the whole surface of the neutron star at $L_a> 0.34
L_{\rm ed}$ as it is well seen in Fig.\,1.

\section*{CONCLUSIONS}
\noindent
The previously obtained representative sample of thermonuclear
X-ray bursts detected by the JEM-X and IBIS/ISGRI telescopes
onboard the INTEGRAL observatory allowed the dependence of the
rate of X-ray bursts generated by the bursters on their
persistent X-ray luminosity (accretion rate indicator) to be
reproduced. It follows from its analysis:

\begin{enumerate}
\item in the range of high luminosities, $4\times
  10^{36}\ \mbox{erg s}^{-1}\la L\,(3-20\ \mbox{keV}) \la
  2\times 10^{37}\ \mbox{erg s}^{-1}$ the observed rate exceeds
  by several times that rate expected in the model of complete
  burning during a burst of the fuel that fell to the neutron
  star surface upon accretion after the previous burst;

\item the observed excess of the burst rate can be explained in
  terms of the model of a spreading layer of accreting matter
  over the neutron star surface (Inogamov and Sunyaev 1999),
  according to which the accreting matter is accumulated on the
  stellar surface in two or three separate ring zones, each of
  which can be responsible for its burst;

\item the cases where a burst in one of the zones initiates the
  propagation of a flame over the stellar surface capable of
  igniting the thermonuclear fuel in other zones that were
  described in detail by Grebenev and Chelovekov (2017) (see
  also Keek et al. 2010) are realized quite rarely, in a few
  percents of all similar events;

\item the burst rate in the model of a spreading layer can rise
  also due to the fact that the matter spreading over the
  surface of the neutron star occupies rather a limited area
  (two high-latitude ring zones) compared to the surface of the
  entire star. Correspondingly, the critical surface density
  necessary to begin the explosion is reached earlier than in
  the spherically symmetric model. The area of the ring zones is
  defined by their width and latitude and, finally, by the
  accretion rate.
\end{enumerate}

\noindent
We will investigate these and other consequences
of the model of a spreading layer in our subsequent
papers.
\vspace{3mm}

\section*{ACKNOWLEDGMENTS}
\noindent
This work is based on the long-term observations performed by
the \mbox{INTEGRAL} international gamma-ray astrophysics
observatory and retrieved via its Russian and European
Scientific Data Centers. We are grateful to the Russian Science
Foundation for financial support (grant no. 14-22-00271).


\begin{flushright}
{\sl Translated by V. Astakhov\/}
\end{flushright}
\end{document}